\documentclass{llncs}
\usepackage{tikz}
\usetikzlibrary{shapes.multipart}
\usetikzlibrary{automata}
\usetikzlibrary{positioning}
\usepgflibrary{shapes.gates.logic.US}
\usetikzlibrary{shapes.gates.logic.US}
\usetikzlibrary{shadows}
\usetikzlibrary{arrows}
\usetikzlibrary{trees}

 \usepackage[T1]{fontenc} \usepackage{tgtermes}

\title{Explaining Violation Traces with Finite State Natural Language Generation Models}
\author{Gordon J. Pace \and Michael Rosner}
\institute{University of Malta\\
\texttt{gordon.pace|mike.rosner@um.edu.mt}
}
\date{}

\begin{document}
\maketitle

\begin{abstract}
An essential element of any verification technique is that of identifying and communicating to the user, system behaviour which leads to a deviation from the expected behaviour. Such behaviours are typically made available as long traces of system actions which  would benefit from a natural language explanation of the trace  and especially in the context of business logic level specifications. In this paper we present  a natural language generation model which can be used to explain such traces. A key idea is that the explanation language is a CNL that is, formally speaking, regular language susceptible transformations that can be expressed with finite state machinery. At the same time it admits various forms of abstraction and simplification which contribute to the naturalness of explanations that are communicated to the user. 
\end{abstract}

\section{Introduction}
The growth in size and complexity of computer systems has been accompanied by an increase in importance given to the application of verification techniques, attempting to avoid or at least mitigate problems arising due to errors in the system design and implementation. Given a specification of how the system should behave (or, dually, of what the system should not do), techniques ranging from testing to runtime verification and model checking attempt to answer the question of whether or not the system is correct. One common issue with all these techniques, is that a negative answer is useless unless accompanied by a trace showing how the system may perform leading to a violation of the expected behaviour. 

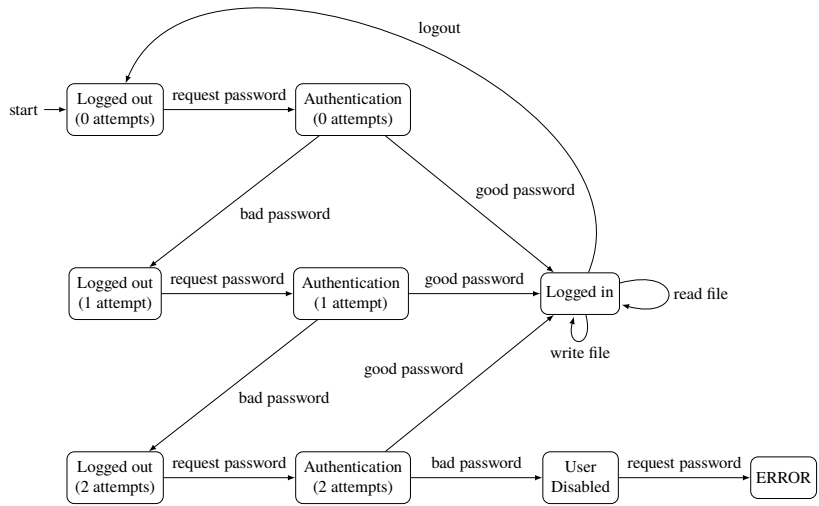
\begin{figure}
\begin{center}
\scalebox{0.7}{\makebox{
\begin{tikzpicture}[
  >=latex,
  node distance=2.5cm,
  auto,
  every state/.style={rectangle,rounded corners,draw,font=\footnotesize},
  every node/.style={font=\footnotesize},
  bend angle=90,
  ]
\node[state,initial] (LO0)
   { \begin{tabular}{c}
     Logged out\\(0 attempts)
     \end{tabular} };
\node[state] (LO1) [below=of LO0]
   { \begin{tabular}{c}
     Logged out\\(1 attempt)
     \end{tabular} };
\node[state] (LO2) [below=of LO1]
   { \begin{tabular}{c}
     Logged out\\(2 attempts)
     \end{tabular} };
\node[state] (A0) [right=of LO0]
   { \begin{tabular}{c}
     Authentication\\(0 attempts)
     \end{tabular} };
\node[state] (A1) [right=of LO1]
   { \begin{tabular}{c}
     Authentication\\(1 attempt)
     \end{tabular} };
\node[state] (A2) [right=of LO2]
   { \begin{tabular}{c}
     Authentication\\(2 attempts)
     \end{tabular} };

\node[state] (UD) [right=of A2]
   { \begin{tabular}{c}
     User\\Disabled
     \end{tabular} };
\node[state] (FAIL) [right=of UD] {ERROR};
\node[state] (LI) [right=of A1] {Logged in};

\path[->]
  (LO0) edge node {request password} (A0)
  (LO1) edge node {request password} (A1)
  (LO2) edge node {request password} (A2)

  (A0)  edge node {good password} (LI)
        edge node {bad password}  (LO1)
  (A1)  edge node {good password} (LI)
        edge node {bad password}  (LO2)
  (A2)  edge node {good password} (LI)
        edge node {bad password}  (UD)

  (LI)  edge [loop right] node {read file} ()
        edge [loop below] node {write file} ()
        edge [bend right] node [swap] {logout} (LO0)

  (UD) edge node {request password} (FAIL)
;
\end{tikzpicture}
}}
\end{center}
\caption{\label{fig:specification} An automaton-based specification} 
\end{figure}

Consider, for example, the specification of a system which allows user to log in, as shown in Figure \ref{fig:specification},  which states that ``after three consecutive failed user authentications, users should not be allowed to attempt another login''. A testing or runtime verification tool may deduce that the system may perform a long sequence of events which lead to a violation. Although techniques have been developed to shorten such counter-examples \cite{shortcounterexamples}, such traces may be rather long, and using them to understand the circumstances in which the system failed to work as expected may not always be straightforward.

In the case of implementation-level properties and traces, tools such as debuggers and simulators may enable the processing of long traces by developers to understand the nature of the bug, but in the case of higher-level specifications, giving business-logic level properties, such traces may need to be processed by management personnel. For example, a fraud expert may be developing fraud rules to try to match against the behaviour of known black-listed users, and may want to understand why a trace showing the behaviour of such a user does to trigger a rule he may have just set up. In such cases, a natural language explanation of such a trace would help the expert to understand better what is going wrong and why.

In this paper, we present the use of finite state natural language generation (NLG) models to explain violation traces. We assume that the basis of the controlled natural language used to describe such behaviour is given by the person writing the specification, by articulating how the actions can be described using a natural language, and how they can be abstracted into more understandable explanations. We present a stepwise refinement of the process, explaining how a more natural feel to the  generated controlled natural language text can be given using finite state techniques.

Although the work we present is still exploratory, we believe that the approach can be generalised to work on more complex systems, and it can give insight into how far out the limits of finite state NLG techniques can be pushed.

\section{The Roles of NLG and CNL}

In this section we illustrate a solution to the problem of generating reasonably natural explanations from sequences of the above type in a computationally efficient way. The two critical ingredients are (i) NLG, which, in a general sense,  provides a set of techniques for generating text flexibly given an abstract non-linguistic representation of semantic content,  and (ii) CNLs which, in a nutshell, are natural languages with a designer element --- natural in the sense that they can be understood by native speakers of the ``parent'' language and designed to to be simpler than that language from some computational perspective such as translation to logic or, as in this paper, NLG. An excellent survey and classification scheme for CNLs appears in Kuhn~\cite{Kuhn:2014}

The final output of NLG is clearly natural language of some kind. The nature of the process that produces that output is somewhat less clear, in that there are still many approaches though most research in the area is consistent with the assumption that that it includes at least the stages of content planning (deciding what to say), content packaging (packaging the content into sentence-sized messages), and surface realisation (constructing individual sentences). These three stages are linked together in a pipeline, according to the architecture proposed by Reiter and Dale~\cite{Reiter-Dale:2000}. 

The complexity of NLG arises from fact that the input content severely underdetermines the surface realisation and that there are few guiding principles available to narrow the realisation choices. Consequently, the process is even more nondeterministic than the inverse process of of natural language {\it understanding} where at least it is possible to appeal to common sense when attempting to choose amongst competing interpretations. With NLG, some dimensions of complexity must be sacrificed  for the computation to be feasible.

In this paper, the sacrifice comes down to a choice concerning two sets of languages: (i) that which expresses the content, which we will call the $C$ language, and  (ii) a sequence of languages in which explanations are realised that we will call $E_1$, $E_2$, etc. $C$ is a form of semantic representation language, whilst $E_1, E_2, \ldots E_n$ are CNLs in Kuhn's sense.

In both cases, we assume that both $C$ and $E_i$ languages are {\it regular languages} in the formal sense. This has a number of advantages: the computational properties of such languages are well understood, and we know that algorithms for parsing and generation of are of relatively low complexity. Additionally, we can express linguistic processes as relations over such languages that can be computed by finite-state {\it transducers}. Elementary transductions can be composed together to carry out complex linguistic processing tasks. Using techniques originally advocated for morphological analysis by Beesley and Kattunen~\cite{Beesley-Karttunen:2003} we can envisage a complex NLG process as a series of finite state transductions combined together under relational composition, thus opening up the possibility of describing the synthesis of an explanation to efficient, finite-state machinery.

Of course, the restriction to regular languages imposes certain limitations upon what content can possibly be expressed in C, and may also impact the naturalness of the violation description expressed in $E$. However, these are empirical issues that will not be tackled in this paper

We are not the first to have used simplified languages in the attempt to reduce the complexity of natural language processing. In the domain of NLG, Wilcock~\cite{Wilcock:2001} proposed \emph{``the use of XML-based tools to implement existing well-known approaches to NLG''}. Power~\cite{Power:2012} uses finite-state representations for expressing descriptions of OWL-LITE sentences. It is of course in the area of computational morphology where finite state methods are best known. 

The main contribution of the paper is to substantiate and present the hypothesis that according to the choice of $C$ and $E$, it is possible to realise a family of efficient NLG systems that are based on steadfastly finite-state technology.

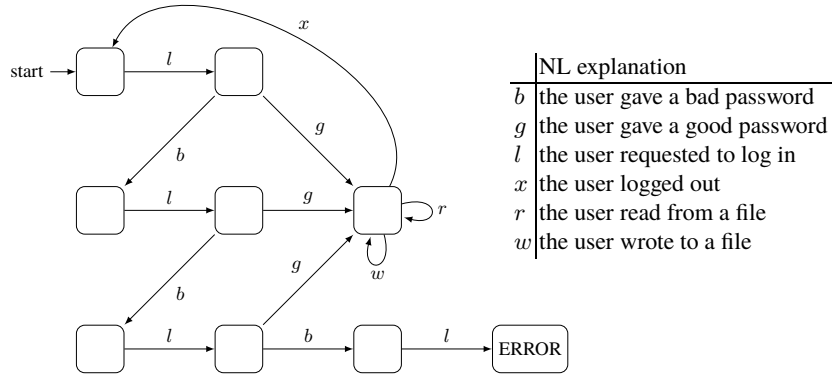
\begin{figure}
\centering
\begin{minipage}{0.55\textwidth}
\scalebox{0.8}{\makebox{
\begin{tikzpicture}[
  >=latex,
  node distance=1.5cm,
  auto,
  every state/.style={rectangle,rounded corners,draw,font=\footnotesize},
  every node/.style={font=\footnotesize},
  bend angle=90,
  ]
\node[state,initial] (LO0)
   { \begin{tabular}{c}
     
     \end{tabular} };
\node[state] (LO1) [below=of LO0]
   { \begin{tabular}{c}
     
     \end{tabular} };
\node[state] (LO2) [below=of LO1]
   { \begin{tabular}{c}
     
     \end{tabular} };
\node[state] (A0) [right=of LO0]
   { \begin{tabular}{c}
     
     \end{tabular} };
\node[state] (A1) [right=of LO1]
   { \begin{tabular}{c}
     
     \end{tabular} };
\node[state] (A2) [right=of LO2]
   { \begin{tabular}{c}
     
     \end{tabular} };

\node[state] (UD) [right=of A2]
   { \begin{tabular}{c}
     
     \end{tabular} };
\node[state] (FAIL) [right=of UD] {ERROR};
\node[state] (LI) [right=of A1] {};

\path[->]
  (LO0) edge node {$l$} (A0)
  (LO1) edge node {$l$} (A1)
  (LO2) edge node {$l$} (A2)

  (A0)  edge node {$g$} (LI)
        edge node {$b$}  (LO1)
  (A1)  edge node {$g$} (LI)
        edge node {$b$}  (LO2)
  (A2)  edge node {$g$} (LI)
        edge node {$b$}  (UD)

  (LI)  edge [loop right] node {$r$} ()
        edge [loop below] node {$w$} ()
        edge [bend right] node [swap] {$x$} (LO0)

  (UD) edge node {$l$} (FAIL)
;
\end{tikzpicture}
}}
\end{minipage}
\begin{minipage}{0.35\textwidth}
\small
\begin{tabular}{l|l}
 & NL explanation\\\hline
$b$ & the user gave a bad password \\
$g$ & the user gave a good password \\
$l$ & the user requested to log in \\
$x$ & the user logged out \\
$r$ & the user read from a file \\
$w$ & the user wrote to a file 
\end{tabular}
\normalsize
\end{minipage}
\caption{The specification augmented with NL explanations}
\label{fig:taggedspecification}
\end{figure}

\section{Languages}

In what follows we first present the $C$ language and then a sequence of $E$ languages, progressively adding features to attain a more natural explanation of the trace. As we shall investigate in more detail in Section \ref{s:fsg}, at each stage we use further information to obtain more natural generated text. 

\subsection{The $C$ Language}
We assume that the basic specification of the $C$ language is given by the automaton shown in Figure \ref{fig:taggedspecification}.
The following trace is a sentence
$$lgrxlblgwwxlgrwxlgxlblblbl$$
Note that although the automaton itself is not necessary for the explanations that ensue, it could in principle be used to check which trace prefix leads to an error state to allow for an explanation when such a state is reached.

Next we turn to the series of $E$ languages. Since these are all CNLs we will refer to them as CNL0, CNL1, CNL2 and CNL3 respectively. All four languages are similar insofar as they all talk about the same underlying, domain-specific world of states and actions, and they all finite state. At the same time they are somewhat different linguistically.


\subsection{CNL0}

Sentences of the CNL0 language are very simple declarative sentences of the kind that we typically associate with simple predicate-argument structures. In the example shown in Figure \ref{fig:CNL0} here, each sentence has a subject, a verb, and possibly a direct object.

\begin{figure}
\centerline{\begin{tabular}{|lcl|}\hline \ \hspace{1ex}\ &&\ \hspace{1ex}\ \\ &\begin{minipage}{0.8\textwidth}
\emph{The user requested to log in. The user gave a good password. The user read from a file. The user logged out. The user requested to log in. The user gave a bad password. The user requested to log in. The user gave a correct password. The user wrote to a file. The user wrote to a file. The user logged out. The user requested to log in. The user gave a correct password. The user read from a file. The user wrote to a file. The user logged out. The user requested to log in. The user gave a good password. The user logged out. The user requested to log in. The user gave a bad password. The user requested to log in. The user gave a bad password. The user requested to log in. The user gave a bad password. The user requested to log in, which should not have been allowed.}
\end{minipage}&\\&&\\\hline\end{tabular}}
\caption{A na\"ive explanation of the trace: CNL0}
\label{fig:CNL0}
\end{figure}

In this paper, the mapping between the $C$ language and CLN0 is given extensionally by means of a {\it lexicon} that connects the individual transition names with a sentence with a simple and fixed syntactic structure. The lexicon itself is expressed as a finite state transducer, as described in Section \ref{naivegen}. For more complex systems such an approach might not be practical, and a solution could then be to {\it derive} the sentence associated with each transition from more fundamental properties of the underlying machine. 

CNL0 provides for a somewhat na\"ive explanation of traces using the explanations provided by the domain expert directly. 

\subsection{CNL1}

Next we turn to CNL1 which offers some improvements. The main feature of CNL1 is that it is a sequence of paragraphs, where each paragraph is simply a sequence of CNL0 sentences, as shown in Figure \ref{fig:CNL1}

\begin{figure}
\centerline{\begin{tabular}{|lcl|}\hline \ \hspace{1ex}\ &&\ \hspace{1ex}\ \\ &\begin{minipage}{0.8\textwidth}
\begin{enumerate}
\item The user requested to log in. The user gave a good password. The user read from a file. The user logged out.
\item The user requested to log. The user gave a bad password. 
\item The user attempted to log in. The user gave a good password. The user wrote to a file. The user wrote to a file. The user logged out. 
\item The user requested to log in. The user gave a good password. The user read from a file. The user wrote to a file. The user logged out. 
\item The user requested to log in. The user gave a good password. The user logged out. 
\item The user requested to log in. The user gave a bad password.
\item The user requested to log in. The user gave a bad password.
\item The user requested to log in. The user gave a bad password.
\item The user requested to log in, which should not have been allowed.
\end{enumerate}
\end{minipage}&\\&&\\\hline\end{tabular}}
\caption{A grouped explanation: CNL1}
\label{fig:CNL1}
\end{figure}

There are two consequences to this slightly richer structure. One is that it provides the skeleton upon which to hang the numbered steps. This is a presentation issue that arguably increases the naturalness and improves comprehension. The other is that it gives a structural identity to each paragraph that could be exploited in order to attribute certain semantic properties to the associated sequence of actions. For example, we have the notion of {\it correctness} which has the potential to figure in explanations. Nevertheless, this property is not actually exploited in CNL1.

\subsection{CNL2}

The main novelty in CNL2, (see Figure \ref{fig:CNL2}) in contrast to CNL1, is the use of {\it aggregation} to reduce each multi-sentence paragraph to a single, more complex sentence. This is a technique which is used for removing redundancy (see Dalianis and Hovy~\cite{Dalianis-Hovy:1993}), yielding texts that are more fluid, more acceptable and generally less prone to being misunderstood by human readers than CNL1-style descriptions.

\begin{figure}
\centerline{\begin{tabular}{|lcl|}\hline \ \hspace{1ex}\ &&\ \hspace{1ex}\ \\ &\begin{minipage}{0.8\textwidth}
\begin{enumerate}
\item The user requested to log in, gave a correct password and after reading from a file logged out. 
\item The user requested to log in, and gave a bad password. 
\item After a log in request the user gave a correct password and wrote twice to a file before logging out. 
\item The user requested to log in, gave a correct password, read from a file, wrote to a file and then logged out. 
\item After requesting a log in, the user gave a good password and logged out. 
\item The user requested to log in, gave a bad password, requested again to log in, gave another bad password and after requesting to log in, gave another bad password. 
\item Finally, the user made a request to log in, which should not have been allowed.
\end{enumerate}
\end{minipage}&\\&&\\\hline\end{tabular}}

\caption{A better grouped explanation: CNL2}
\label{fig:CNL2}
\end{figure}

The linguistic renderings resulting from aggregation in CNL2 include:

\begin{enumerate}
\item Punctuation other than full stops
\item Temporal connectives (``after'', ``then'', ``finally'')
\item The use of contrastive conjunctions like ``but"
\item Collective terms (``twice")
\end{enumerate}

\subsection{CNL3}

Finally CNL3 (see Figure \ref{fig:CNL3}) is considerably more complex, because it not only contains further aggregation but also {\it summarisation}.

\begin{figure}
\centerline{\begin{tabular}{|lcl|}\hline \ \hspace{1ex}\ &&\ \hspace{1ex}\ \\ &\begin{minipage}{0.8\textwidth}
\emph{The user logged in a number of times, interspersed by sequences of one or two bad logins, after which she unsuccessfully attempted to log in 3 times. The user then made another request to log in, which should not have been allowed.}
\end{minipage}&\\&&\\\hline\end{tabular}}

\caption{A natural explanation: CNL3}
\label{fig:CNL3}
\end{figure}

In this example, there are only two sentences. The first sentence not only aggregates the first six sentences, but it also omits some of the information (for example, the the user read from a file, that the user logged out etc.). It also includes the use of certain phrases whose correct interpretation, as mentioned earlier, requires consideration of the context of occurrence as well as use of adverbs (``she unsuccessfully attempted") and the use of more complex tenses (``should not have been allowed").


\section{Finite State Generation\label{s:fsg}}

%
%
%
In this section we will look into using finite state CNLs for NLG. This is based on finite state techniques as embodied in {\tt xfst} (Beesley and Karttunen~\cite{Beesley-Karttunen:2003}) that has already been used extensively in several other areas of language processing such as computational morphology and light parsing. {\tt xfst} provides a language for the description of complex transducers together with a compiler and a user interface for running and testing transducers. Our aim is to better understand the tradeoffs involved between producing reasonably natural explanations from traces and the use of the efficient computational machinery described here.

\subsection{Na\"ive Generation: CNL0}
\label{naivegen}

Just as in Figure \ref{fig:taggedspecification}, our starting point is a regular input language C defined as follows

\begin{verbatim}
define SIGMA b|l|g|x|r|w;
define C SIGMA*;
\end{verbatim}

{\tt SIGMA} is the alphabet of the original FSA and the entire generation mechanism accepts inputs that are arbitrary strings over this alphabet. Strings containing illegal characters yield the empty string and hence, no output.
 
CNL0 can be obtained more or less directly via a dictionary which links symbols in {\tt SIGMA} to simple declarative sentences, as follows\footnote{Some of the syntactically more obscure aspects of this definition have been omitted for the sake of clarity.}:

\begin{verbatim}
define SP " ";
define USR {the SP user};
define DICT  b->  [{user} SP {gives} SP {bad}  SP {password}],
             l->  [{user} SP {requests} SP {login}],
             g->  [{user} SP {gives} SP {good} SP {password}], 
             x->  [{user} SP {logs} SP {out}],
             r->  [{user} SP {reads} SP {from} SP {a} SP {file}],
             w->  [{user} SP {writes} SP {to} SP {a} SP {file}];


define CNL0 C .o. DICT;
\end{verbatim}

The first line defines the space character, and the second the symbol USR. The third defines the dictionary DICT which is implemented as finite state transducer that maps from the individual action symbols to primitive sentences, all of which have the same basic structure.
The input sequence is represented as a string
\begin{verbatim}
define input {lgrxlblgwwxlgrwxlgxlblblbl};
\end{verbatim}

To get the output we compose {\tt CNL0} with {\tt input} using the expression 
({\tt [input .o. CNL0]}), extract the {\it lower} side of the relation with the $l$ operator ({\tt [input .o. CNL0].l}). The problem with the generated output is that there are no separators between the sentences. The solution is to compose the input with a transducer {\tt sentencesep} that inserts a separator. 
\begin{verbatim}
input .o. sentencesep .o. CNL0
\end{verbatim}
This turns the input into the following string: 
\begin{verbatim}
        l.g.r.x.l.b.l.g.w.w.x.l.g.r.w.x.l.g.x.l.b.l.b.l.b.l.
\end{verbatim}
Such a string can be made to yield exactly the sentences of CNL0 by arranging for the mapping of the fullstops to insert a space. This is just another transducer that is composed into the pipeline.
The result of this process is exactly the text shown in Figure \ref{fig:CNL0}.

\subsection{Adding Structural Information: CNL1}

At a simplest level, we can specify how the explanation may be split into an enumerated sequence of paragraphs, aiding the comprehension of the trace explanation. Consider being given the following list of subtrace specifications using regular explanations:

\begin{description}
\item[Correct login session:] $lg(r+w)^*x$.
\item[Sequence of incorrect login requests:] $(lb)^*$.
\end{description}

In CNL1, the main feature is that we will use this information to group text. We will assume that the  following paragraph definitions are supplied:

\begin{verbatim}
define correct l g [r | w]* x;
define incorrect [l b];
define group1 correct @-> ... %|, incorrect @-> ... %|;
\end{verbatim}

\noindent The {\tt group1} definition includes a piece of {\tt xfst} notation that causes a vertical bar to be inserted just after whatever matched the left hand side of the rule, yielding  
\begin{verbatim}
lgrx|lb|lgwwx|lgrwx|lgx|lb|lb|lb|l
\end{verbatim}

As shown earlier, we can when applied to the input, where the vertical bar is used to delimit paragraphs.

\begin{figure}
\begin{verbatim}
       l.g.r.x.|l.b.|l.g.w.w.x.|l.g.r.w.x.|l.g.x.|l.b.|l.b.|l.b.|l.
\end{verbatim}
\caption{CNL1 representation just prior to lexicalisation\label{fig:cnl1input}}
\end{figure}

Composing this with an augmented version of CNL0 that also handles the paragraph breaks yields exactly the paragraph structure of the CNL1 rendering shown in Figure \ref{fig:CNL1}. An inherent limitation of this approach is that it is impossible to produce a finite-state transducer that will output a numbering scheme for arbitrary numbers of paragraphs. Our solution is to postprocess the output, and generate, for instance HTML or \LaTeX\ output which will handle the enumeration as required..

\subsection{Adding Aggregation: CNL2}

We can now move on to CNL2. This involves several intermediate stages which are diagrammed below:

\begin{verbatim}
A: l.g.r.x.|l.b.|l.g.w.w.x.|l.g.r.w.x.|l.g.x.|l.b.|l.b.|l.b.|l.
B: l,g,r,x.|l,b.|l,g,w,w,x.|l,g,r,w,x.|l,g,x.|l,b.|l,b.|l,b.|l.|
C: aggregation1
D: aggregation2
\end{verbatim}

\texttt{A} is as shown in Figure \ref{fig:cnl1input}. We must now prepare for aggregation by first replacing all but the paragraph-final fullstops with commas. Because the transducer that achieves this uses the paragraph marker to identify the final fullstop, we must first insert that final paragraph marker as as shown in \texttt{B}. The next two phases of aggregation are best explained with the following example: we wish to transform ``the user requested to login. the user gave a good password. the user logged out.'' to the more natural ``the user requested login, gave a good password, and logged out''. The first phase removes the subject (i.e. the phrase `` the user'') of all sentences but reinstates the same subject at the beginning of the paragraph. The second inserts an ``and" just before the final verb phrase of each aggregated sentence. In this way we are able to achieve paragraph 2 of the CNL2 example as shown in Figure Similar, surface-oriented techniques can be used to obtain the other paragraphs in Figure \ref{fig:CNL2}. Specifically, we have composed rules for inserting the words ``after", ``then", ``twice", ``another", ``finally" and ``and". However, space limitations prevent us from describing these in full.

\subsection{Adding Abstraction: CNL3\label{s:addingabstraction}}
We note that certain sequences of actions can be combined into a simpler explanation, abstracting away (possibly) irrelevant detail, thus aiding comprehension. For instance, consider the following rules, consisting of (i) a regular expression matching a collection of subtraces which may be explained more concisely; and (ii) a natural language explanation which may replace the detailed text one would obtain from the whole subtrace:  

\begin{description}
\item[Consecutive correct login sessions:] 
$(lg(r+w)^*x)^n$ explained as \emph{``The user successfully logged in $n$ times''}.
\item[Consecutive correct failed login attempts:] 
$(lb)^n$ explained as \emph{``The user unsuccessfully attempted to log in $n$ times''}.
\item[Correct login sessions interspersed with occasional incorrect one:] 
$((lg(r+w)^*x)*lb(lg(r+w)^*x)+)^*$ explained as \emph{``The user successfully logged in a number of times, with one off bad logins in between''}.
\item[Correct login sessions interspersed with occasional incorrect one or two:]
$((lg(r+w)^*x)*(lb+lblb)(lg(r+w)^*x))^*$ explained as \emph{``The user logged in a number of times, interspersed by sequences of one or two bad logins''}.
\end{description}

Note that {\tt xfst} allows regular expressions that are parametrised for the number of times a repeated expression matches. For example, the statement 
\begin{verbatim}
define success3 [l g [r | w]* x]^3; 
\end{verbatim}

\noindent achieves the first definition above and associates it with the multicharacter symbol {\tt success3}. This can be added to the dictionary {\tt DICT} and associated with the string in much the same way as the strings associated with transitions, as shown above.

We will assume that these rules will be applied using a maximal length strategy --- we prefer a longer match, and in case of a tie, the first rule specified is applied. {\tt xfst} allows the user to choose between longest and shortest match strategies. Using appropriate {\tt xfst} rules would result in the description given in Figure \ref{fig:CNL3}.

\subsection{Adding Contextuality: CNL4\label{s:addingcontextuality}}

To further enrich the generation explanations, we can extend the approach used in the previous section for CNL3, to allow for actions to  be described using different terms in different contexts. For example, a \emph{logout} action when logged in may be described as \emph{`the user logged out'}, while a \emph{logout} occuring while the user is already logged out would better be described as \emph{`the user attempted to log out'}. We can use techniques similar to the ones presented in the previous section, using regular expressions to specify contexts in which an action will be described in a particular manner. 

Consider the specification below, in which each action and natural language description pair is accompanied by two regular expressions which have to match with the part of the trace immediately preceding and following the action for that description to be used\footnote{We use the notation $\overline{a}$ to signify any single symbol except for $a$.}:

\medskip 
\centerline{\begin{tabular}{c|c|c|l}
\ \ Action\ \ & Pre & Post & \ \ CNL rendering\\\hline
$x$ & \ \ $l\,\overline{x}^*$\ \  & -- & \ \ user logs out \\
\cline{2-4}
    & \multicolumn{2}{l|}{\ \ \emph{otherwise}\ \ }  & \ \ user attempts to log out \\ \hline
$l$  
    & -- & $b$ & \ \ the user attempts to log in\\\cline{2-4}
    & \multicolumn{2}{l|}{\ \ \emph{otherwise}\ \ }  & \ \ the user logs in \\ 
\hline
\end{tabular}}

\medskip\noindent This technique can be further extended and refined to deal with repetition of actions as shown below with repeated logins:

\medskip 
\centerline{\begin{tabular}{c|c|c|l}
\ \ Action\ \  & Pre & \ \ Post\ \  & \ \ CNL rendering\\\hline
$l$  
    & \ \ $l\,b\,\overline{l}^*$\ \  & $b$ & \ \ user attempts to log in again\\\cline{2-4}
    & -- & $b$ & \ \ user attempts to log in\\\cline{2-4}
    & \ \ $l\,\overline{b}\,\overline{l}^*$\ \  & -- & \ \ user logs in again\\\cline{2-4}
    & \multicolumn{2}{l|}{\ \ \emph{otherwise}\ \ }  & \ \ user logs in \\ 
\hline
\end{tabular}}

\medskip\noindent It is interesting to see how far this approach can be pushed and generalised to allow for the generation of more natural sounding text from the input traces.

\section{Discussion and Conclusions}

In this paper we have presented preliminary results illustrating how finite state approaches can be used generate controlled natural language explanations of traces. Although there is still much to be done, the results are promising and it is planned that we use such an approach to allow for the specification of natural language explanations to be used in the runtime verification tool \textsc{Larva} \cite{larva}.

Two problems underlying our task are: (i) the discovery of subsequences that are interesting for the domain in question and (ii) how to turn an interesting subsequence into a natural-sounding explanation. In this paper we have provided somewhat \emph{ad hoc} solutions to both these problems. While one can use profiling techniques to discover interesting, or frequently occurring subsequences, clearly there needs to be a strong human input in identifying which of these sequences should be used to abstract and explain traces more effectively. On the other hand, we see that many of the \emph{ad hoc} solutions adopted to make explanations more natural-sounding may be generalised to work on a wide-range of situations. We envisage that the person building the specification may add hints as to how to improve the explanation, such as the tables shown in Section \ref{s:addingabstraction} to improve abstraction and the ones given in Section \ref{s:addingcontextuality} to add contextuality. 

Given that, essentially, we are using regular grammars to specify our natural language generator, the generalisation process to  reduce human input while generating more natural-sounding text is bound to hit a limit. It is of interest to us, however, to investigate how  far these approaches can be taken without resorting to more sophisticated techniques usually applied to language generation.


\bibliographystyle{alpha}
\bibliography{refs}

\end{document}